# Model for Pressure Induced Deformations in Carbon Nanotube Materials

*Shuchi Gupta, Keya Dharamvir and V.K. Jindal*
*Physics Department, Panjab University, Chandigarh 160 014, India*

*We report the results of a model calculation for studying the effects of hydrostatic pressure on a bunch of carbon nanotubes. At pressures that we work with, the deformation in axial direction comes out to be negligibly small. We find that hydrostatic pressure is an ideal probe to study the radial deformations of the nanotubes. The nanotubes are considered to be flexible, identified by a flattening of cylinders under pressure through a parameter f. We use the 6-exponential and Brenner potentials to account for inter and intra-tube interactions respectively. We calculate the total energy of the deformed tubes in bunches. The free energy thus calculated enables us to calculate phase changes at various pressures. From our calculations, we find the phase transformation to occur at about 5GPa.*

**Introduction**:

Since their discovery in 1991, carbon nanotubes (CNT) have hogged much attention from various research communities. Their high aspect ratio, large tensile strength, the ability to exist in either metallic or semiconducting forms and extreme flexibility make them promising candidates as high strength fibers and various novel nanometer scale electronic and mechanical devices. A CNT can be envisaged as a graphene sheet rolled up into a seamless cylinder to form a macromolecule. They exist as single nanotubes of various kinds as well as materials of these in forms of bunches of single wall or multiwall formations. In a bundle consisting of identical nanotubes, they are arranged in a 2-D hexagonal close packing (forming 'nanoropes'), interacting via weak Vander Waals (VdW) type attractive forces, shown in fig. 1a.

Usually, the cylindrical tubes are assumed to be circular in cross section. However several theoretical models predict them to be deformed either elliptically or having facets due to the VdW forces between the neighboring tubes in a bundle, shown in fig 1b and 1c. These deformations are prominent for larger diameter tubes or when an external strain is applied perpendicular to the long axis of the tubes. Even an isolated tube assumes a flattened or collapsed structure whose extent depends largely on the diameter of the tube [1]. Further, it is harder to distort a multiwalled nanotube (MWNT) than a single walled nanotube (SWNT) [2].

Since the deformation of the cross section of the tube may quite affect various properties, it is of prime importance to study the behavior of SWNT bundles under pressure. There have been various theoretical and experimental evidences available in literature, indicating the pressure driven transformations in SWNTs. Chesnokov[3] et al. conducted the experiment upto a pressure of about 27kbar and observed a large and reversible volume loss of the SWNT bundles. Several Raman spectroscopy studies have also been reported; e.g. Venkateswaran et al.[4] made investigations on the pressure dependence of the Raman active radial and tangential breathing modes of SWNT bundles and attributed the disappearance of radial mode intensity beyond 1.5 GPa to the faceting of nanotubes under pressure. Peters et al [5 studied Raman shifts and reported a structural phase transition at a pressure of about 1.7 GPa. Recent measurements by Tang et al [6] suggest that the tubes of diameter of about 14Å may be slightly polygonized even at zero pressure although this polygonization is more prominent for higher pressures and obtained a structural distortion at about 1.5GPa as a result of diminishing of trigonal lattice, which is reversible upto a pressure of 4GPa. Lopez et al [7] observed the faceting of tubes of about 17Å diameter from HRTEM image. They also did MD simulations and found that the equilibrium configuration of the lattice corresponds to circular tubes; however the lattice of faceted tubes is very close in energy. Rols [8] performed neutron diffraction studies upto 50kbar and found that shape deformation dominates the compression process around 20kbar pressure. However, X-ray studies conducted by Sharma et al [9] showed the vanishing of diffraction line, thus indicating a phase transition ~ 10GPa which is not related to uniform flattening and/or uniform faceting of tubes but is due to loss of triangular lattice and this lattice reappears on unloading of pressure from ~ 13 GPa. Therefore, a complete understanding of these pressure dependent transitions is still required. In this paper, we report calculations showing the reversible deformation of tubes from circular to hexagonal cross section with increase in pressure.

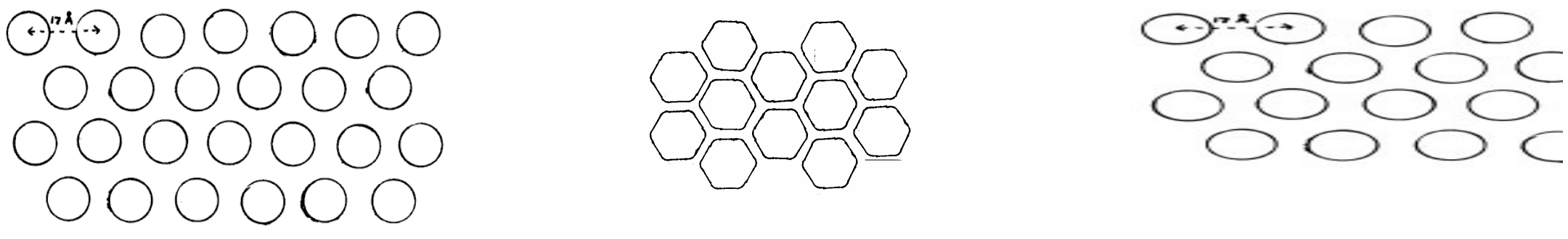

(a) (b) (c)

**Fig. 1**: Cross section through bundle of nanotubes: (a) undistorted tube cross section , (b) faceted tubes, (c) elliptically deformed tubes.

## Model and Calculations:

We study the properties of a nanotube rope in which all the tubes are identical and are of infinite length. The interaction between two carbon atoms on the same tube is modelled by an effective short range potential of Tersoff-Brenner form [10]. For C-atoms on the different tubes, a 6-exp potential [11] is considered.

The short range potential, which describes the covalent bonding was given by Brenner. Its parameters were given by Tersoff to explain diversified forms of carbon, including diamond and graphite. However some of these parameters have been modified by us so as to get a better fit for the bond lengths and cohesive energy of graphite. The potential energy between the atoms $i$ and $j$ on the same tube separated by a distance $r_{ij}$ is of the form:

$$V(r_{ij}) = f_c(r_{ij})[f_R(r_{ij}) + b_{ij} f_A(r_{ij})] \qquad (1)$$

where

$$f_R(r) = Ae^{-\lambda_1 r} \quad ; f_A(r) = -Be^{-\lambda_2 r} \qquad (1a)$$

$f_c$ is a cut off function which is simply taken as:

$$f_c(r) = \begin{cases} 1, & r < (R-D), \\ \frac{1}{2} - \frac{1}{2}\sin[\frac{1}{2}\pi(r-R)/D], & (R-D) < r < (R+D), \\ 0, & r > (R+D), \end{cases}$$

This form of cutoff, which goes from 1 to 0 in a small range around $R$, is continuous and has a derivative for all $r$. $R$ is chosen to include only the first-neighbour shell. The function $b_{ij}$ implicitly includes the bond order and must depend on local atomic environment. All deviations from a simple pair potential are ascribed to the dependence of $b_{ij}$ upon the local atomic environment. Specifically, the bonding strength $b_{ij}$ for the pair $ij$ should be a monotonically decreasing function of the coordination of atoms $i$ and $j$. $b_{ij}$ has the following form:

$$b_{ij} = \frac{1}{(1+\beta^n \varsigma_{ij}^n)^{\frac{1}{2n}}} \quad , \qquad (1b)$$

where, $$\varsigma_{ij} = \sum_k f_c(r_{ij}) g(\theta_{ijk}) \exp[\lambda_3^3 (r_{ij} - r_{ik})^3] \qquad (1c)$$

with $$g(\theta) = 1 + \frac{c^2}{d^2} - \frac{c^2}{[d^2 + (h-\cos\theta)^2]} . \qquad (1d)$$

In eq. (1c) the summation over $k$ runs over neighbours of $i$ leaving out $j$; and $\theta_{ijk}$ is the angle between bonds $ij$ and $ik$.

Most of the parameters used in this potential are taken to be the same as those given by Tersoff [10] except for a few which are modified by us in order to get a better fit to the bond length and cohesive energy of graphite. These parameters are tabulated in Table I.

**Bending Energy of a SWNT:**

Each tube is an elastic sheet bent into a cylindrical form. On bending the graphene sheet to form a tube, the angles between neighbouring bonds change. So the bonding strength $b_{ij}$ also changes. Using the Tersoff-Brenner potential with modified parameters, bending energy for SWNT of the "armchair" variety with various radii R are calculated (shown in Table II). We consider a rolled up graphene sheet consisting of a certain number of hexagonal rings along its circumference. The coordinates of C-atoms on such a cylinder are known (they are generated using those of graphene). The total bond energy for this configuration, using the Brenner potential (eq. 1) is calculated. The coordinates are adjusted till the total energy obtained reaches a minimum value. This is repeated for various diameters of tube. We find that these calculations are in close agreement with the QMD calculations performed by Adams et al [12]. Bending energy is defined as the difference in the bond energies of a tube compared to the total bond energy of same area of a plane sheet of graphene.

**Table I:** The short range potential parameters

| | A (eV) | B (eV) | $\lambda_1$ (Å$^{-1}$) | $\lambda_2$ (Å$^{-1}$) | R (Å) | D (Å) | β ($10^{-7}$) | n | c ($10^{-4}$) | d | h |
|---|---|---|---|---|---|---|---|---|---|---|---|
| Brenner-Tersoff | 1393.6 | 346.7 | 3.4879 | 2.2119 | 1.8 | 0.3 | 1.5724 | .72751 | 3.8049 | 4.3484 | -.57058 |
| Ours | 1380 | 349.4 | 3.5679 | 2.2724 | 1.8 | 0.3 | 1.5724 | .72751 | 3.8049 | 4.3484 | -.57058 |

**Table II:** Bending energy for tubes of different radii.

| Radius(A$^o$) | E (eV/atom) | E(QMD)[12] |
|---|---|---|
| 3.39 | .183 | .182 |
| 4.07 | .126 | .126 |
| 5.43 | .070 | .071 |
| 6.78 | .044 | .045 |

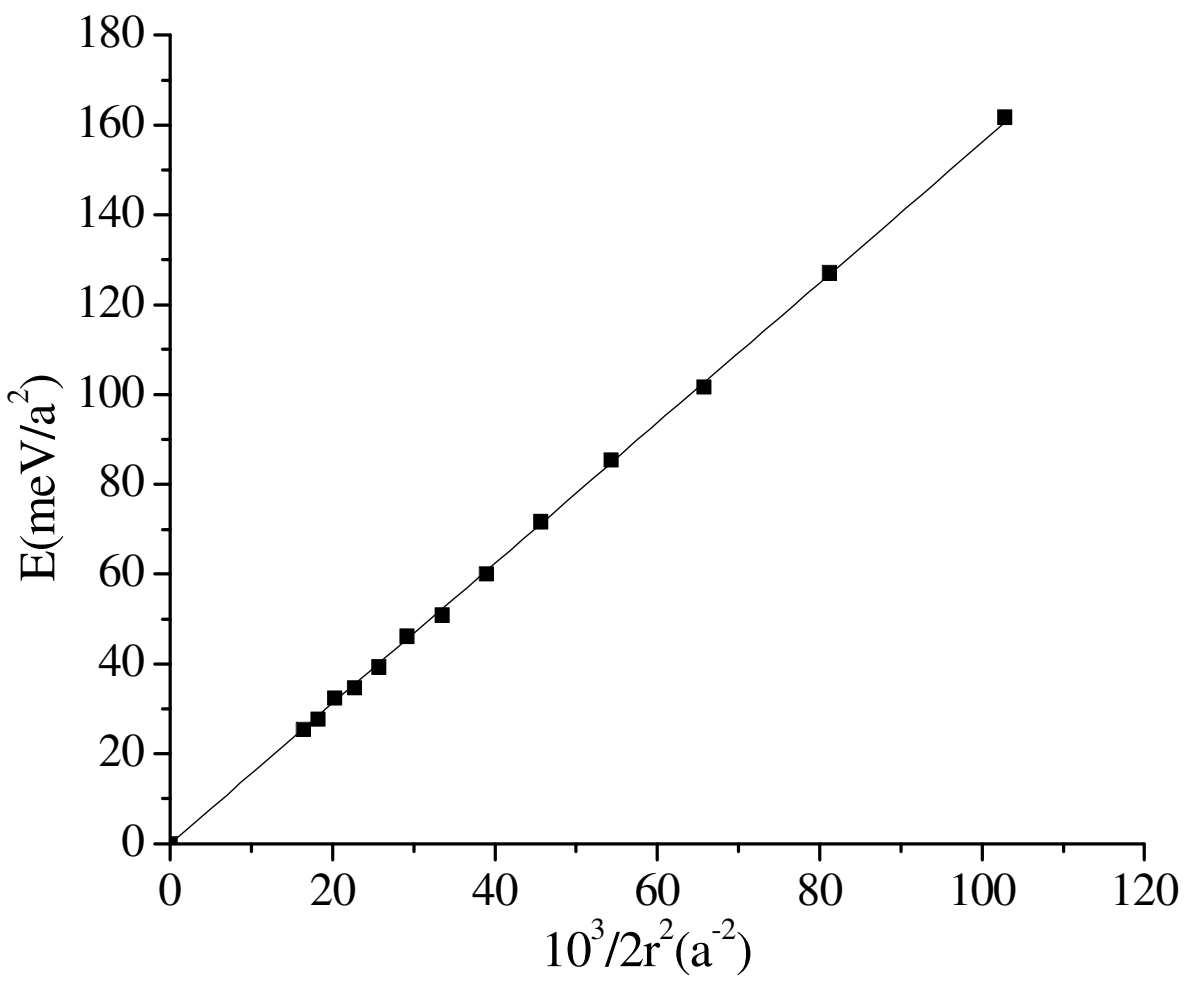

**Fig. 2.** Bending energy as a function of radius of curvature.

Fig. 2 is a plot of bending energy per unit area of tube surface as a function of inverse square radius of the tube. It follows that when a piece of graphene sheet is bent to form part of a cylinder of radius r, the energy spent is proportional to 1/r. Thus the graph between the bending energy per unit area and $1/r^2$ is a straight line as shown. This gives, for the total bending energy associated with the tube,

$$E_{bending} = \frac{\pi c_o L}{r} \qquad (2)$$

where $c_o$ = 1.54 eV and L=length of the tube.

**Effect of the pressure applied radially on the tube**:

The bond length in each bucky tube may be considered rigid. Bond- bending, on the other hand, may be achieved by much lower stress. Therefore, in what follows, we shall consider bending or distortion of tube walls but not bond stretching. Consequently, the tube cross section may change shape, but density of C-atoms per unit area of wall remains same.

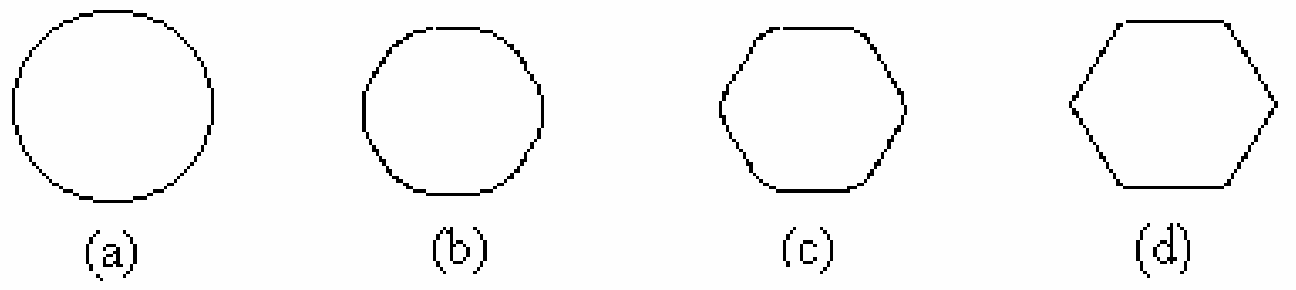

**Fig.5a:** Cross sections for tubes of various distortion, with factor f = (a) 0.0, (b) 0.3, (c) 0.6 and (d)0.9

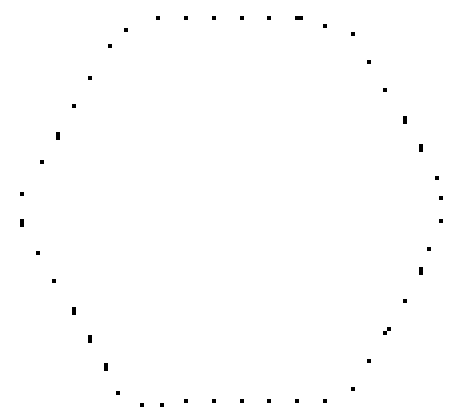

**Fig. 5b:** Cross section of a faceted tube. Each dot represents a rod of C-atoms. Larger the number of dots, closer the model to continuum limit.

As the applied hydrostatic pressure on a bunch is increased, the tubes move closer together, moving against Van der Waals forces. When two neighbouring tubes are sufficiently close and press against each other, they may flatten at the region of contact. This flattening increases the area of contact, whereby a larger number of atoms have come closer, lowering the energy via VdW interaction. However, this also increases the curvature at the corners. Fig. 5 shows the expected cross section of each tube once such a flattening has taken place, taking into account the fact that each tube is surrounded by six neighbours, hence six-sided faceting. The increase in curvature results in the mechanical energy of bending being raised. These two processes compete in energy. In order to study this, we model the distorted (or faceted) tube with the help of a distortion factor $f$ . A fraction f of the total area of a tube is in six flat portions and the rest comprises of six rounded corners. The cross section of such tubes corresponding to different values of $f$ are shown in fig. 5a. The excess energy of bending of such a distorted tube over one with circular cross section is given by [15], using eq. (2),

$$U_{bending} = \pi c_o L(\frac{1}{\rho} - \frac{1}{r}) \qquad (13)$$

where $r$ is the radius of the undistorted (circular cross section) tube and $\rho$ is the radius of curvature at the corner after distortion (faceting). $\rho$ depends on the deformation parameter as $\rho = (1 - f)r$ . $L$ is the length of the tube under consideration. We will see later that $f$ plays the role of order parameter for a faceting phase transition.

We consider parallel identical tubes with a given $f$ , arranged on a close packed 2-D lattice (fig. 1b). In the continuum limit, each tube is supposed to comprise of a large number, N, of evenly distributed thin parallel rods or lines of C-atoms parallel to the axis (fig. 5b). The linear density of C-atoms on each rod depends on the number of rods being used, since areal density of C-atoms on a tube wall is a constant. To find the total interaction energy of this lattice, we sum over interactions between all rod-rod pairs in the system, except intratube interactions, i.e.,

$$U_{tot} = \frac{1}{2} \sum_{i,j}{}' V_{ij} \qquad (14)$$

where prime excludes $ij$ pair belonging to same tube. The interaction $V_{ij}$ is obtained by integrating the 6-exp potential over two parallel lines separated by a distance $r$

$$U(r) = -\frac{A}{r^6} + Be^{-\alpha r} \tag{15}$$

where the interaction parameters $A$, $B$ and α are 358 kcal mol$^{-1}$Å$^6$, 42000 kcal mol$^{-1}$ and 3.58 Å$^{-1}$ respectively, as provided by Kitaigorodsky[13] and integrating this, we get

$$U_{ij} \equiv U(r_{ij}) = \left[ -\frac{3\pi}{8}\frac{A}{r_{ij}^5} + 2Br_{ij}K_1(\alpha r_{ij}) \right]\lambda^2 L \tag{16}$$

where $L$ is the length of each tube, infinitely long in the limit; $\lambda$ = linear density of C-atoms on each rod; $K_1$ is a modified Bessel function; $r_{ij}$ = distance between the two parallel rods. Each rod is, however, not continuous and is made up of atoms as shown on line ABC in fig 4. But since the distortion along lengthwise direction is negligible, we can take it to be continuous. We can calculate the interaction energy between two single walled nanotubes by assuming a smeared out continuous tube model having uniform distribution of carbon atoms on their surfaces and integrating C-C interactions over this distribution. Such a model has successfully reproduced the bulk and lattice properties of $C_{60}$ and $C_{70}$ solids in the past[14]. Performing this integral, we find the tube-tube interaction per unit length as:

$$U_{tube-tube}(r) = -U_{attractive}(r) + U_{repulsive}(r) \tag{17}$$

where

$$U_{attractive}(r) = \frac{8}{3}\pi^3\sigma^2 A \frac{R^2}{r^5} \sum_{n=0}^{\infty} \left\{ (n+\frac{1}{2})(n+\frac{3}{2})^{2n}C_n \right\} \left( \frac{R}{2r} \right)^{2n} \tag{17a}$$

and

$$U_{repulsive}(r) = 8\pi^2\sigma^2 B\{RI_o(\alpha r)\}^2 rK_1(\alpha r)^{2n} \left[ 1 - \frac{2R}{r}\frac{I_1(\alpha r)}{I_o(\alpha r)}\frac{K_o(\alpha r)}{K_1(\alpha r)} \right] \tag{17b}$$

with $R$ = radius of a nano-tube, $r$ = distance between the axes of the two tubes which are parallel to each other and σ = number density of C-atoms on its surface. $I_o$, $I_1$, $K_o$ and $K_1$ are appropriate Bessel functions. ${}^{2n}C_n$ are the Bionomial coefficients. This form, though compact, is restricted to tubes with cylindrical cross sections. Therefore we have not used it here where, in addition to bunches with circular cross sections, we have to consider tubes with faceted cross sections too.

The lattice energy of such a configuration is found by numerically summing over interactions of rods of a given tube with those of all its neighbours. We find that 48 rods per tube and a lattice distance of one neighbour are sufficient for numerical accuracy.

To the lattice energy thus obtained, we add the bending energy of the distorted tubes according to eq. (13). We thus obtain $U(f, r)$, the total energy of the lattice as a function of lattice size, given by $R$, and the distortion parameter $f$. For a given $f$, equilibrium lattice constant $R_o$, is obtained by plotting $U$ against $R$ and identifying coordinates where the minimum of energy occurs. Please see fig. 5, where the three minimum $R_o(0)$, $R_o(0.3)$ and $R_o(0.5)$ are shown. These same curves are used to find pressure, as follows.

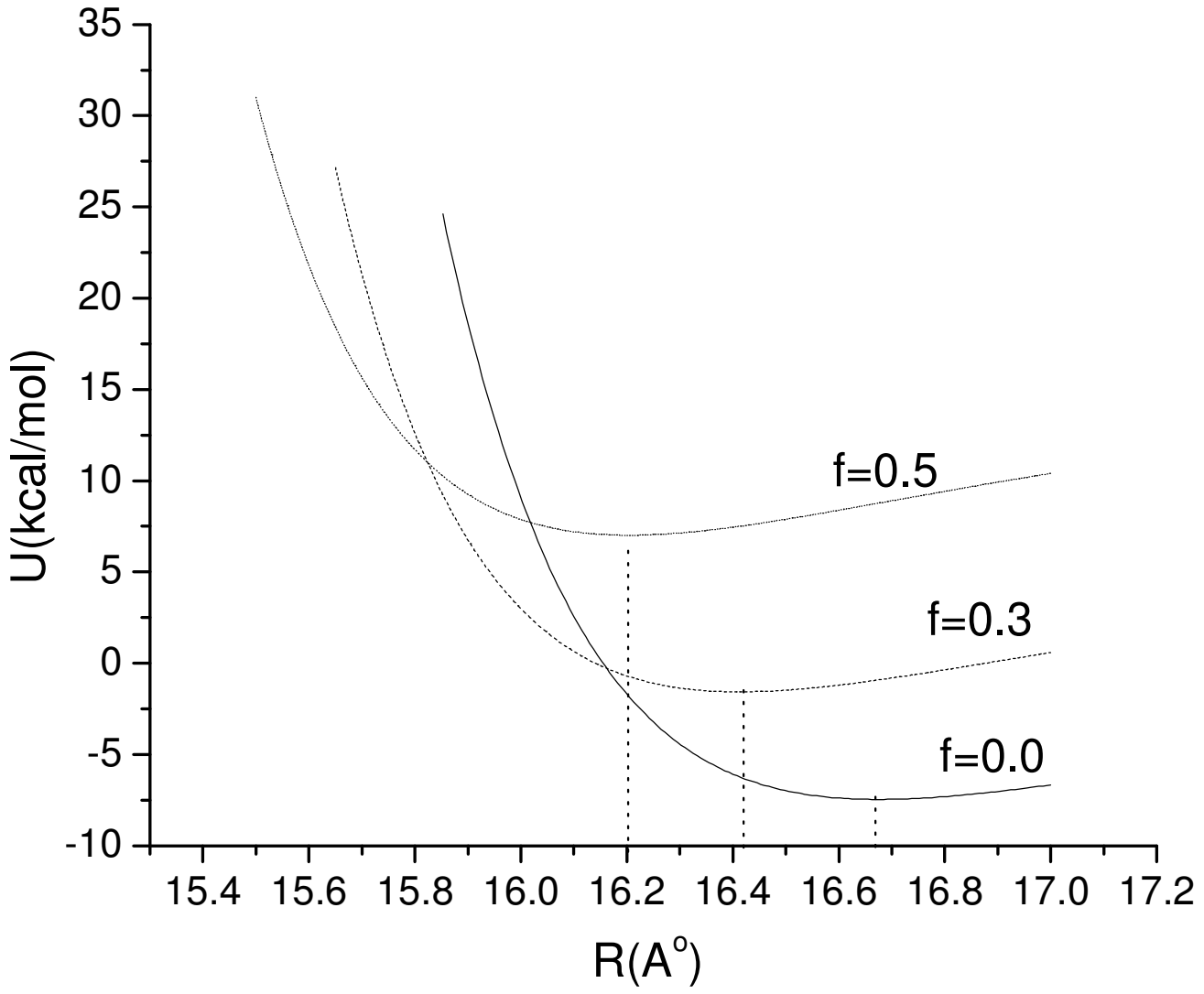

**Fig.5:** The interaction energy between two bucky-tubes. The three curves are for circular cross section, slightly distorted (faceted) and dominantly faceted tube cross sections

At zero pressure, the system would be in equilibrium with $R = R_o$. however when compressed, it would acquire a state with $R$ such that

$$U = U_o + P\Delta V \qquad (18)$$

such that

$$\Delta V = \sqrt{3}(R_o^2 - R^2) \qquad (19)$$

where we have used $V = \sqrt{3}R^2$, $V$ being volume per tube per unit length.

Thus from the given curve (fig. 5) for $U$ vs $R$, we get $U$ vs $P$. This $U$ (eq. (18)) is the Helmholtz free energy at T=0 by definition. At a given $P$, when one compares free energy $U(f)$ for all $f$, the stable system is given by minimum $U(f)$, i.e., system adopts that $f$ for which U is minimum for that pressure. This yields $f$ vs $P$ curve, Fig.6. This shows that for $P < P_c$, $f$ =0, gives stable configuration; whereas for $P > P_c$, $f$ rises as $P$ increases. At t=0, $P_c$ is the initial point where order parameter $f$ begins to rise above zero. Once we have $f$ vs $P$ for a particular $P$, we can pick up R corresponding to appropriate $f$. Thus we have $R^2$ (or $V$) vs $P$. this is shown in Fig. 7. The $P$-$V$ curve has a kink at $P_c$.

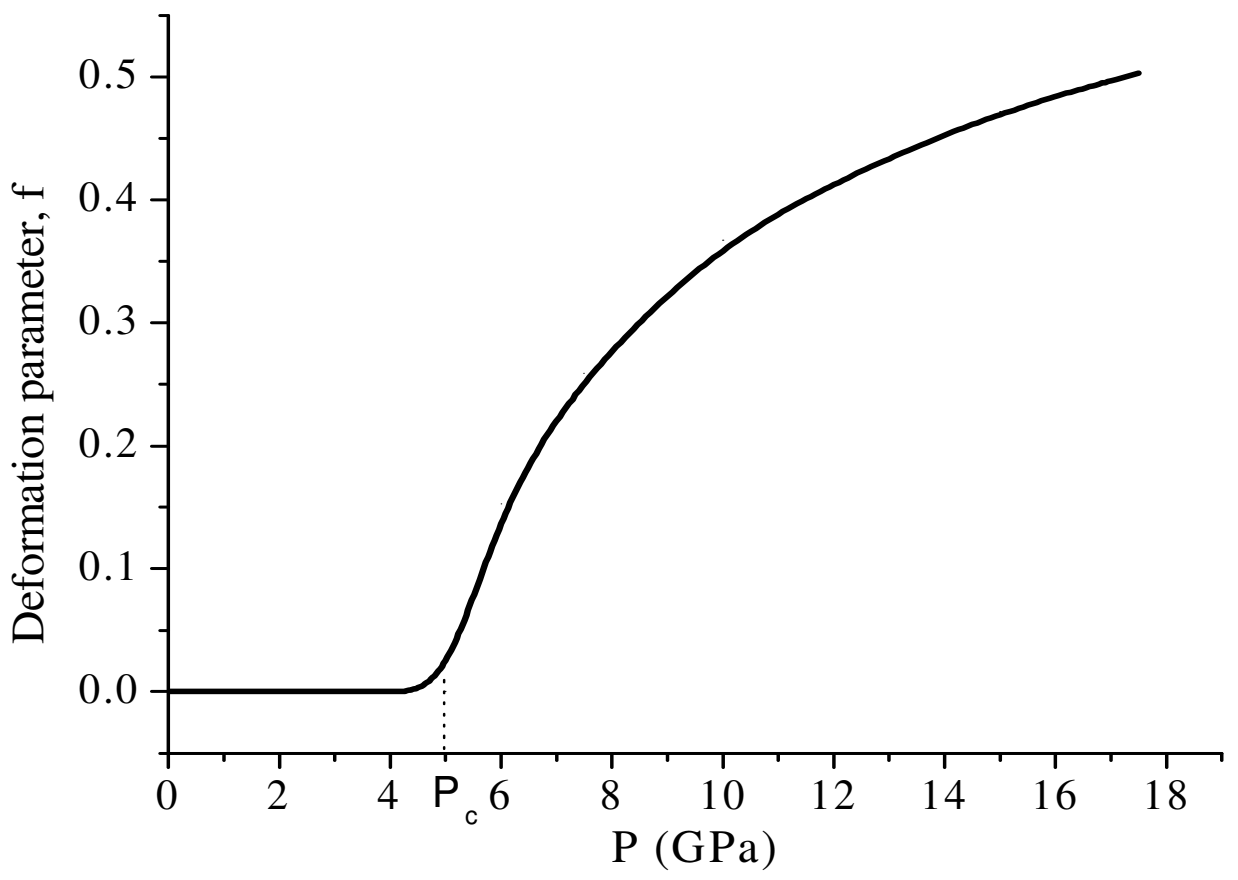

**Fig. 6:** Stable tube cross section at various pressures

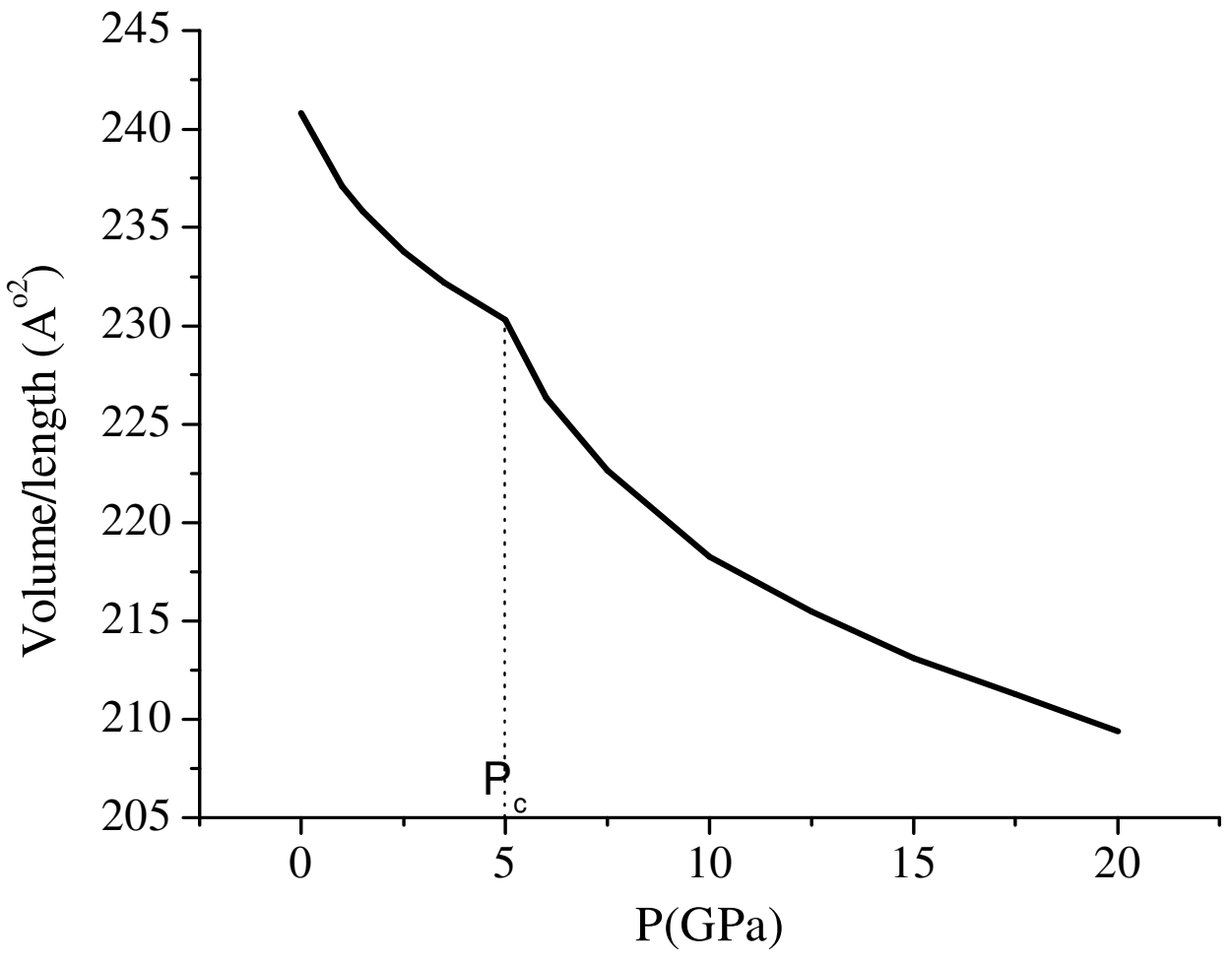

**Fig 7**: Variation of volume with pressure

Similarly, the compressibility, K, is obtained from the p-V curve as:

$$K = -\frac{1}{V}\frac{\partial V}{\partial P} \tag{20}$$

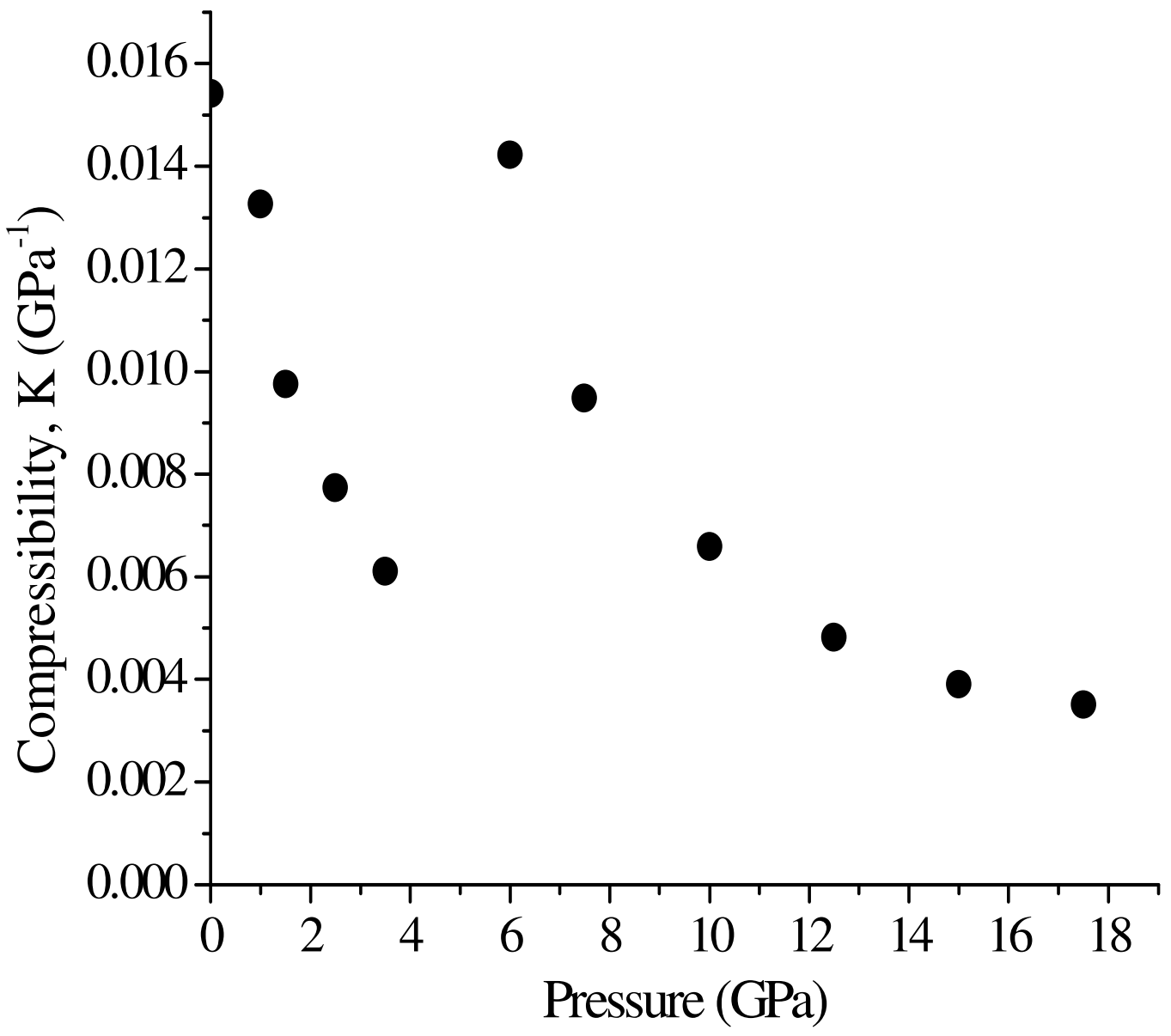

**Fig 8**: Variation of compressibility with pressure. The solid points represent the points at which calculation was made.

**Conclusions:**

The discontinuity in the compressibility curve indicates a structural phase transition at a pressure of about 5Gpa since the order parameter $f$ changes continuously across $P_c$, though its first derivative is discontinuous, we identify it as a second order phase transition. This result is not very compatible with the values obtained experimentally, but the work presented here gives a good qualitative understanding of the pressure induced phase transitions. From our model, we found that about 34Å diameter tubes show faceting even at ambient conditions, however Tersoff [15] found this for 25Å diameter tubes and Lopez et al. [7] observed this faceting for 17Å tubes. This indicates that the continuum model, used by us for making calculations, make the tubes a little harder, thereby explaining the transition appearing at a higher pressure than observed experimentally. Therefore, to reproduce better quantitative results, discrete atomic positions of the tubes should be considered. Phase transition also depends on the value of elastic bending constant. A lower value of this constant makes the discontinuity in the compressibility curve to come at a lower pressure. E.g. $c_0$=15 kcal/mole corresponds to a transition at a pressure=1.7GPa.

**Acknowledgements:**
The financial assistance provided by CSIR, New Delhi in the form of fellowship is greatly acknowledged by SG.